# GAUSSIAN ARMA MODELS IN THE `ts.extend` PACKAGE

BEN O'NEILL[*], *Australian National University*[**]

WRITTEN 26 SEPTEMBER 2021

**Abstract**

This paper introduces and describes the `R` package `ts.extend`, which adds probability functions for stationary Gaussian ARMA models and some related utility functions for time-series. We show how to use the package to compute the density and distributions functions for models in this class, and generate random vectors from this model. The package allows the user to use marginal or conditional models using a simple syntax for conditioning variables and marginalised elements. This allows users to simulate time-series vectors from any stationary Gaussian ARMA model, even if some elements are conditional values or omitted values. We also show how to use the package to compute the spectral intensity of a time-series vector and implement the permutation-spectrum test for a time-series vector to detect the presence of a periodic signal.

GAUSSIAN ARMA MODEL; R; SPECTRAL INTENSITY; PERMUTATION-SPECTRUM TEST.

Gaussian auto-regressive moving average (ARMA) models are a standard class of models used in time-series analysis. These models were initially derived out of a combination of auto-regressive models (Yule 1926) and moving-average models (Slutsky 1937), but they were fleshed out further with empirical tools and popularised in Box and Jenkins (1970). Statistical literature on ARMA models is extensive and there are a number of textbooks on the topic. Two notable texts are Brockwell and Davis (1991) and Fuller (1996), which both set out the theory in detail. There have also been a number of extensions to ARMA models in the literature, but they are regarded as a core model for the analysis of time-series; ARMA models and their variants and extensions are surveyed extensively in Holan, Lund and Davis (2000).

It is well-known from Wold's theorem that any covariance-stationary time series process can be represented as a deterministic component plus a moving-average process of infinite order (Wold 1954). Stationary ARMA models can also be represented as moving-average processes of infinite order. Although the former class of model is larger than the class of ARMA models, a great benefit of the latter is that they are parsimonious models that are "dense" with respect to the class of all short-memory covariance-stationary time series. This means that the ARMA form allows representation of an extremely broad class of time-series processes up to arbitrary accuracy. This makes them a broad class of model with a relatively simple form.

---

[*] E-mail address: ben.oneill@anu.edu.au; ben.oneill@hotmail.com.
[**] Research School of Population Health, Australian National University, Canberra ACT 0200.

ARMA models are known to relate closely to signal analysis in the frequency domain, with attention given to the relationship between statistical time-series models defined in the time domain and their properties in the frequency domain (see e.g., Priestley 1981, Yaglom 1987). While ARMA are widey used in time-series analysis, there has been some concern that high-order models can overfit data and lead to poor out-of-sample performance. Makridakis and Hibon (1998) give a summary and analysis of literature examining forecasting performance of ARMA and ARIMA models, and they find that simple low-order models tend to perform as well as —or better than— complex high-order models for out-of-sample forecasting. In any case, due to their dense coverage of the class of covariance-stationary models, ARMA models remain a core part of time-series analysis, and are widely used both in their own right, and as a building block for more complex time-series models.

There is a substantial existing cache of **R** packages and functions for time-series analysis, and these packages have been updated and developed over time to give a substantial suite of useful functions. Some core packages for time-series analysis in **R** include the **forecast** package (Hyndman *et al* 2021), the **zoo** package (Zeileis *et al* 2021), the **tseries** package (Trapletti *et al* 2020), the **fts** package (Armstrong 2018), the **xts** package (Ryan *et al* 2020), the **tsutils** package (Kourentzes *et al* 2020) and the **TSstudio** package (Krispin 2020). Other packages such as **lubridate** (Spinu et al 2021), **timeDate** (Wuertz *et al* 2018) and **chron** (James *et al* 2020) are also of potential assistance for dealing with time and date variables.

In this paper we introduce the **ts.extend** package in **R**, containing probability and statistical functions for stationary Gaussian ARMA models and some related utility functions for time-series analysis. Our package fills an important gap in the existing computational packages for time-series analysis, by allowing users to obtain exact computations of density functions and cumulative distribution functions for stationary Gaussian ARMA models, allowing users to simulate time-series vectors from these models, and giving some additional facilities for testing for periodic signals. The ARMA functions in our package accommodate both marginal and conditional distributions for these models, allowing users to obtain probabilities or simulations that are conditional on stipulated values occurring at particular times. The inclusion of facilities for conditional analysis is particularly useful, since it is non-trivial to simulate from these conditional distributions. This gives users a highly general facility to simulate data and conduct probabilistic analysis for stationary Gaussian ARMA models.

# 1. Stationary ARMA models

Time-series analysis often uses auto-regressive moving average (ARMA) models, which allow both an auto-regressive and a moving-average component in the model. These processes are usually described by the recursive equation for the observable time-series process $\{y_t | t \in \mathbb{Z}\}$ based on is an underlying noise process $\{\varepsilon_t | t \in \mathbb{Z}\}$. An ARMA$(p, q)$ a process with mean $\mu$ is described by the **recursive model equation**:

$$y_t = \mu + \sum_{i=1}^{p} \phi_i (y_{t-i} - \mu) + \varepsilon_t + \sum_{i=1}^{q} \theta_i \varepsilon_{t-i},$$

where the values $\phi_1, \ldots, \phi_p$ are the auto-regression coefficients and the values $\theta_1, \ldots, \theta_q$ are the moving average coefficients. Usually the underlying noise process is composed of independent and identically distributed error terms with zero mean and error variance $\sigma^2$.

The ARMA model form is usually written in a more succinct form using the backshift operator $\mathfrak{B}$, which operates on values in either the observable time-series or the underlying error process to move them backward one unit in time.[1] To facilitate analysis we define the **auto-regressive characteristic polynomial** and **moving-average characteristic polynomial** respectively by:

$$\phi(x) = 1 - \phi_1 x - \cdots - \phi_p x^p \qquad \theta(x) = 1 + \theta_1 x + \cdots + \theta_q x^q.$$

Using the backshift operator we can then write the recursive equation for the model as:

$$\phi(\mathfrak{B})(y_t - \mu) = \theta(\mathfrak{B}) \varepsilon_t.$$

Use of the polynomial form allows decomposition using the roots of the polynomials. Letting $r_1, \ldots, r_p$ be the roots of the auto-regressive characteristic polynomial and $s_1, \ldots, s_q$ be the roots of the moving-average characteristic polynomial, these polynomials can be written as products of affine functions as follows:

$$\phi(x) = \prod_{i=1}^{p} \left(1 - \frac{x}{r_i}\right) \qquad \theta(x) = \prod_{i=1}^{q} \left(1 - \frac{x}{s_i}\right).$$

If there are any common roots of the polynomials then the ARMA model can be reduced by cancelling these common terms. We therefore assume that this has already occurred, and so there are no common roots in the polynomials.

---

[1] This operator gives $\mathfrak{B} y_t = y_{t-1}$ and $\mathfrak{B} \varepsilon_t = \varepsilon_{t-1}$. It operates on constants to give $\mathfrak{B} k = k$. For more detailed information on the use of the backshift operator in ARMA models, see Kasparis (2016).

The vast majority of models for time-series analysis stipulate model forms that are stationary, and this requires us to impose additional requirements on the model parameters. In particular, it is well-known that the auto-regressive characteristic polynomial is invertible if all its roots are outside the unit circle (i.e., we have $\min_i |r_i| > 1$). In this case we can use the formula for the sum of a geometric series to obtain the inverse:

$$\frac{1}{\phi(x)} = \prod_{i=1}^{p} \left(1 - \frac{x}{r_i}\right)^{-1} = \prod_{i=1}^{p} \sum_{k=0}^{\infty} \left(\frac{x}{r_i}\right)^k,$$

and we define the **transfer function** $\psi$ as the rational function:

$$\psi(x) \equiv \frac{\theta(x)}{\phi(x)} = \left(1 + \sum_{i=1}^{q} \theta_i x^i\right) \left(\prod_{i=1}^{p} \sum_{k=0}^{\infty} \left(\frac{x}{r_i}\right)^k\right) = \sum_{i=0}^{\infty} \psi_i x^i.$$

(The leading coefficient of this polynomial is $\psi_0 = 1$.) For a stationary ARMA model it can be shown that $\sum_{i=0}^{\infty} |\psi_i| < \infty$ (which also means that $\sum_{i=0}^{\infty} |\psi_i \psi_{i+\ell}| < \infty$) which means that the time-series process has "short memory". The transfer function lets us write the ARMA($p,q$) model in its MA($\infty$) representation:

$$y_t = \mu + \psi(B)\varepsilon_t = \mu + \varepsilon_t + \sum_{i=1}^{\infty} \psi_i \varepsilon_{t-i}.$$

The main advantage of the MA($\infty$) representation of the model is that it allows us to derive the first two moments of the observable time-series. It is standard to assume that the underlying noise process $\{\varepsilon_t | t \in \mathbb{Z}\}$ is a second-order stationary process with zero mean. This means that we have $\mathbb{E}(\varepsilon_t) = 0$ and $\mathbb{Cov}(\varepsilon_t, \varepsilon_{t-\ell}) = \sigma^2 \mathbb{I}(\ell = 0)$. Applying the MA($\infty$) representation then allows us to obtain the mean and covariance for the observable time-series, which is:

$$\mathbb{E}(y_t) = \mu \qquad \mathbb{Cov}(y_t, y_{t-\ell}) = \sigma^2 \gamma(\ell) \qquad \gamma(\ell) \equiv \sum_{i=1}^{\infty} \psi_i \psi_{i+\ell}.$$

This confirms that $\mu$ is the mean parameter of the process. Since the mean is a fixed parameter value and the covariance between any two values of the observable time-series does not depend on time except through the lag $\ell$ between the two values, the time-series $\{y_t | t \in \mathbb{Z}\}$ is second-order stationary. It is worth noting that Wold's theorem shows that any time-series that is mean and covariance stationary can be represented as an MA($\infty$) process on an underlying series of second-order stationary error terms. The latter is a broader case than the ARMA model, and it also includes "long memory" processes. Nevertheless, the ARMA model can approximate any short-memory covariance-stationary process up to arbitrary accuracy.

## 2. The GARMA distribution

It is common to impose further structure on the ARMA model by assuming that the noise terms underlying the process are independent and identically distributed Gaussian random variables (called "white noise"). This gives the additional model assumption:

$$\varepsilon_t \sim \text{IID } N(0, \sigma^2).$$

With this additional assumption on the noise, we may refer to the Gaussian ARMA (GARMA) model. Under the GARMA model the values in the time-series $\{y_t | t \in \mathbb{Z}\}$ follow a multivariate Gaussian distribution with the above mean and covariance values. Since the first two moments of the Gaussian distribution fully define the distribution, the stationary GARMA model leads to a unique distribution for any observable vector of values from the process. This distribution is a particular case of the multivariate normal distribution, but it is useful to demarcate this class of distributions as its own family, which we will call the **GARMA distribution**. The distribution exhibits strong stationarity, which makes it particularly convenient for modelling time-series processes.

We will look at the GARMA distribution for a time-series vector $\boldsymbol{y} = (y_1, \dots, y_m)$ composed of $m$ consecutive elements. We begin by looking at the marginal distribution of the time-series vector, but we will later generalise to look at the conditional distribution where some elements are conditioning values. To facilitate the definition of the GARMA distribution, we define the matrix function:

$$\boldsymbol{\Lambda} \equiv \boldsymbol{\Lambda}(\boldsymbol{\phi}, \boldsymbol{\theta}) \equiv \begin{bmatrix} \gamma(0) & \gamma(1) & \gamma(2) & \cdots & \gamma(m-1) \\ \gamma(1) & \gamma(0) & \gamma(1) & \cdots & \gamma(m-2) \\ \gamma(2) & \gamma(1) & \gamma(0) & \cdots & \gamma(m-3) \\ \vdots & \vdots & \vdots & \ddots & \vdots \\ \gamma(m-1) & \gamma(m-2) & \gamma(m-3) & \cdots & \gamma(0) \end{bmatrix}.$$

Under the GARMA model, the time-series vector $\boldsymbol{y} = (y_1, \dots, y_m)$ has a multivariate Gaussian distribution with mean vector $\mathbb{E}(\boldsymbol{y}) = \mu \boldsymbol{1}$ and variance matrix $\mathbb{V}(\boldsymbol{y}) = \sigma^2 \boldsymbol{\Lambda}(\boldsymbol{\phi}, \boldsymbol{\theta})$. We will define this as the **marginal GARMA distribution**, with density function given by:

$$\begin{aligned} \text{GARMA}(\boldsymbol{y} | \mu, \sigma^2, \boldsymbol{\phi}, \boldsymbol{\theta}) &\equiv N(\boldsymbol{y} | \mu \boldsymbol{1}, \sigma^2 \boldsymbol{\Lambda}(\boldsymbol{\phi}, \boldsymbol{\theta})) \\ &= \frac{1}{(2\pi)^{n/2} \sigma^n |\boldsymbol{\Lambda}(\boldsymbol{\phi}, \boldsymbol{\theta})|^{1/2}} \\ &\quad \times \exp\left( -\frac{1}{2} \cdot \frac{(\boldsymbol{y} - \mu \boldsymbol{1})^\text{T} \boldsymbol{\Lambda}(\boldsymbol{\phi}, \boldsymbol{\theta})^{-1} (\boldsymbol{y} - \mu \boldsymbol{1})}{\sigma^2} \right). \end{aligned}$$

The above form is the marginal distribution of the time-series vector $\mathbf{y} = (y_1, \ldots, y_m)$ from the stationary GARMA model. We can generalise this distribution by looking at the conditional distribution of some elements of the time-series vector conditional on other elements in the vector. To do this, we use an indicator vector $\mathbf{c} = (c_1, \ldots, c_m)$ that indicates which elements are conditioning values in the initial vector. (The marginal case above corresponds to $\mathbf{c} = \mathbf{0}$.) We then decompose $\mathbf{y}$ into an argument vector and conditioning vector given respectively by:

$$\mathbf{y}_\triangledown = (y_i | c_i = 0) \qquad \mathbf{y}_\blacktriangle = (y_i | c_i = 1).$$

To go with this decomposition, we likewise decompose the vector $\mu \mathbf{1}$ and matrix $\mathbf{\Lambda}$ as:[2]

$$\boldsymbol{\mu}_\triangledown = \mu \mathbf{1}_\triangledown \qquad\qquad \boldsymbol{\mu}_\blacktriangle = \mu \mathbf{1}_\blacktriangle,$$

$$\mathbf{\Lambda}_{\triangledown,\triangledown} = [\Lambda_{i,j} | c_i = 0, c_j = 0] \qquad \mathbf{\Lambda}_{\triangledown,\blacktriangle} = [\Lambda_{i,j} | c_i = 0, c_j = 1],$$

$$\mathbf{\Lambda}_{\blacktriangle,\triangledown} = [\Lambda_{i,j} | c_i = 1, c_j = 0] \qquad \mathbf{\Lambda}_{\blacktriangle,\blacktriangle} = [\Lambda_{i,j} | c_i = 1, c_j = 1].$$

Using standard formulae for the conditional moments of the multivariate Gaussian distribution, we have:

$$\boldsymbol{\mu}_{\triangledown|\blacktriangle} \equiv \mathbb{E}(\mathbf{y}_\triangledown | \mathbf{y}_\blacktriangle) = \mu \mathbf{1}_\triangledown + \mathbf{\Lambda}_{\triangledown,\blacktriangle} \mathbf{\Lambda}_{\blacktriangle,\blacktriangle}^{-1} (\mathbf{y}_\blacktriangle - \mu \mathbf{1}_\blacktriangle),$$

$$\mathbf{\Lambda}_{\triangledown|\blacktriangle} \equiv \mathbb{V}(\mathbf{y}_\triangledown | \mathbf{y}_\blacktriangle) = \mathbf{\Lambda}_{\triangledown,\triangledown} - \mathbf{\Lambda}_{\triangledown,\blacktriangle} \mathbf{\Lambda}_{\blacktriangle,\blacktriangle}^{-1} \mathbf{\Lambda}_{\blacktriangle,\triangledown}.$$

This gives us the **conditional GARMA distribution**, with density function given by:

$$\begin{aligned}
\text{GARMA}(\mathbf{y}_\triangledown | \mu, \sigma^2, \boldsymbol{\phi}, \boldsymbol{\theta}, \mathbf{y}_\blacktriangle) &\equiv \text{N}(\mathbf{y}_\triangledown | \boldsymbol{\mu}_{\triangledown|\blacktriangle}, \sigma^2 \mathbf{\Lambda}_{\triangledown|\blacktriangle}) \\
&= \frac{1}{(2\pi)^{n/2} \sigma^n |\mathbf{\Lambda}_{\triangledown|\blacktriangle}|^{1/2}} \\
&\quad \times \exp\left( -\frac{1}{2} \cdot \frac{(\mathbf{y} - \boldsymbol{\mu}_{\triangledown|\blacktriangle})^\text{T} \mathbf{\Lambda}_{\triangledown|\blacktriangle}^{-1} (\mathbf{y} - \boldsymbol{\mu}_{\triangledown|\blacktriangle})}{\sigma^2} \right).
\end{aligned}$$

The above distributional form gives a general conditional form of the GARMA distribution. Although we presented the initial vector $\mathbf{y}$ as one that contains *consecutive* time-series values, we can easily generalise beyond this by allowing values in this vector to be "marginalised out" of the analysis. In that case, we simply remove the marginalised elements out of the mean vector and variance matrix prior to performing the conditioning, so that we now regard both $\mathbf{y}$ and $\mathbf{c}$ as vectors taken over a set of distinct time indices that may not be consecutive.

---

[2] Here we take $\mathbf{1}_\triangledown$ and $\mathbf{1}_\blacktriangle$ to be unit vectors with lengths equal to the respective lengths of $\mathbf{y}_\triangledown$ and $\mathbf{y}_\blacktriangle$. Note also that $\mathbf{\Lambda}_{\blacktriangle,\triangledown} = \mathbf{\Lambda}_{\triangledown,\blacktriangle}^\text{T}$ so the formulation has some slight redundancy.

# 3. Implementing the GARMA distribution in the `ts.extend` package

The `ts.extend` package contains functions to compute the density function and cumulative distribution function for the conditional GARMA distribution, or generate random time-series vectors from this distribution. These are implemented in the `dGARMA`, `pGARMA` and `rGARMA` functions. In addition to these probability functions, the package also includes the `ARMA.acf` function to compute a vector of values from the auto-covariance/auto-correlation function, and the `ARMA.var` function to compute the variance matrix. The main functions in the package and their arguments are shown in Table 1 below. (Note that in addition to the functions for the GARMA distribution the package also contains functions to compute the intensity of a time-series vector and implement a permutation spectrum test for a time-series vector; we discuss these functions in a later section.) The package also includes custom print functions for some of its objects, to print these in a user-friendly output format.

| Function | Purpose | Arguments |
|---|---|---|
| `ARMA.acf` | Auto-covariance/ correlation vector | `n,`<br>`ar = numeric(0), ma = numeric(0),`<br>`corr = FALSE` |
| `ARMA.var` | Variance matrix | `n, condvals = NA,`<br>`ar = numeric(0), ma = numeric(0),`<br>`corr = FALSE` |
| `dGARMA` | Density function | `x, cond = FALSE,`<br>`mean = 0, errorvar = 1,`<br>`ar = numeric(0), ma = numeric(0),`<br>`log = FALSE` |
| `pGARMA` | Cumulative distribution function | `x, cond = FALSE,`<br>`mean = 0, errorvar = 1,`<br>`ar = numeric(0), ma = numeric(0),`<br>`log = FALSE` |
| `rGARMA` | Generate random vectors | `n, m, condvals = NA,`<br>`mean = 0, errorvar = 1,`<br>`ar = numeric(0), ma = numeric(0),`<br>`log = FALSE` |
| `intensity` | Fourier intensity | `x,`<br>`centered/centred = TRUE, scaled = TRUE,`<br>`nyquist = TRUE` |
| `spectrum.test` | Permutation-spectrum test | `x,`<br>`sims = 10^6, progress = TRUE` |

**Table 1:** Main functions in the `ts.extend` package

The probability functions `dGARMA`, `pGARMA` and `rGARMA` in the package will work for any stationary GARMA model. The arguments for each of the probability functions are shown in Table 2 below, along with descriptions of how those arguments are treated in the functions. The stationarity requirement means that the auto-regressive characteristic polynomial for the model must have all its roots outside the unit circle, which imposes requirements on the `ar` input for all of the functions. We do not require the model to be "invertible" and so there is no corresponding requirement imposed on the moving-average characteristic polynomial. The functions are programmed using a "defensive programming" method that checks all the inputs, and so it will give user-friendly error messages if any of their inputs are given non-allowable values. Note that if the inputs `ar` and `ma` are not specified, they use an empty default value corresponding to a process of auto-regressive/moving average degree zero. If neither input is specified then there is no auto-regressive/moving average part, and so the resulting GARMA distribution is composed of IID values from a normal distribution.

| Argument | Type | Description |
|---|---|---|
| `x` | Numeric (vector/matrix) | An input vector or matrix. If the input is a vector it is treated as a single time-series. If the input is a matrix, each row of the matrix is treated as a time-series. |
| `n` | Integer (positive) | An integer specifying the number of time-series vectors to generate. |
| `m` | Integer (positive) | An integer specifying the length of each time-series vector to generate. |
| `cond*` | Logical (vector) | Logical values indicating that a value in the time-series enters as a conditioning variable rather than an input to the density/distribution function. This input must either be a single logical value `FALSE` (indicating that there are no conditioning values) or it must be a logical vector with the same length as the number of time values in the input time-series (and at least one element must be `FALSE` so that there is at least one value that is not a conditioning value). |
| `condvals*` | Numeric (vector) | Numeric conditioning values for generating a random time-series. This input must either be a single value `NA` (indicating that there are no conditioning values) or it must be a numeric vector with length `m`. Each numeric value in the vector is taken as a conditioning value and each `NA` shows that there is no conditioning value for that element (i.e., that element is to be randomly generated). |
| `mean*` | Numeric (scalar) | The mean parameter for the GARMA distribution. |
| `errorvar*` | Numeric (positive scalar) | The error variance parameter for the GARMA distribution. |
| `ar*` | Numeric (vector) | The vector of auto-regressive coefficients for the GARMA distribution. |
| `ma*` | Numeric (vector) | The vector of moving-average coefficients for the GARMA distribution. |
| * Non-mandatory argument; default values are given in the function. | | |

**Table 2:** Arguments for the `dGARMA`, `pGARMA` and `rGARMA` functions

All of these functions allow conditioning and marginalisation of values via a relatively simple syntax. In the case of the `dGARMA` and `pGARMA` functions we input a vector or matrix `x` that holds one or more time-series of the same length. (If a matrix is used as the input, each row is treated as a single time-series.) If we want to marginalise out a value in a time-series vector in `x`, we simply set that value to `NA`. To specify conditioning values we use a logical vector `cond` that is the same length as each time-series vector; a `TRUE` value in this logical vector tells the function to treat the associated input as a conditioning value rather than a value that appears in the argument of the density function.[3] All conditioning values must be given numeric values, so the function will give an error if any of the conditioning values are set to `NA` (i.e., an element cannot be both a conditioning and a marginal value). In the event that all of the values are set to be conditioning values, or there are no non-marginal input values as remaining arguments, the output density/probability is set equal to one and the function gives a warning message.[4]

In the case of the `rGARMA` function we input the desired dimensions `n` and `m`, where `n` is the number of time-series vectors to be generated and `m` is the length of each time-series vector to be generated. The output is a matrix of values of class `'time.series'` with each generated time-series vector appearing as one row. It is not necessary to worry about marginalising out unwanted values with this function, since any marginal values can be generated and then later removed by an appropriate sub-setting command (or simply ignored). For conditioning values we use a numeric vector `condvals` input that is the same length as each time-series vector; an `NA` value in this vector tells the function that the associated input is not a conditioning element (i.e., it is to be generated randomly) and a numeric value is treated as a conditioning value in the distribution.[5] The user can either set the `condvals` vector as a numeric vector with some `NA` elements and some numeric elements, or they can set the input to `NA` if there are no conditioning elements. In the event that all values are conditioning values the function will return outputs that are these conditioning vectors (i.e., there is no randomness in this case).

---

[3] Note that the functions are "vectorised" to allow an input matrix `x` containing multiple time-series. However, the `cond` vector applies to each of these series, so the functions do not allow stipulation of different conditioning values for each time-series in `x`. If you want to use different conditioning values for different time-series vectors you can still do this, but you need to call the functions multiple times.
[4] This "conventional" output value is chosen in order to assist with likelihood calculations; by using an output of one we ensure that the output does not affect products of densities.
[5] Again, this function is "vectorised" to allow the user to generate `m` time-series vectors with a single function call. However, the `condvals` vector applies to each time-series, so the functions do not allow stipulation of different conditioning values for each time-series. If you want to use different conditioning values for different randomly generated time-series vectors you can still do this, but you need to call the functions multiple times.

In the code below we show an example of the analysis of a stationary GARMA(2,2) model using the coefficients $\boldsymbol{\phi} = (\phi_1, \phi_1) = (0.8, -0.2)$ and $\boldsymbol{\theta} = (\theta_1, \theta_2) = (0.6, 0.3)$. (We use the default mean of zero and the default error variance of one.) We first use the **rGARMA** function to generate $n = 16$ random time-series vectors of length $m = 30$ from this model and then we plot these time-series vectors using the inbuilt plotting function for this object. The resulting scatterplot matrix is shown in Figure 1 below; each time-series is shown by a blue line and the background shows the points from all the generated series.

```
#Set the model parameters
AR <- c(0.8, -0.2)
MA <- c(0.6,  0.3)

#Generate and plot random time-series from the GARMA distribution
set.seed(1014745590)
SERIES <- rGARMA(n = 16, m = 30, ar = AR, ma = MA)
plot(SERIES)
```

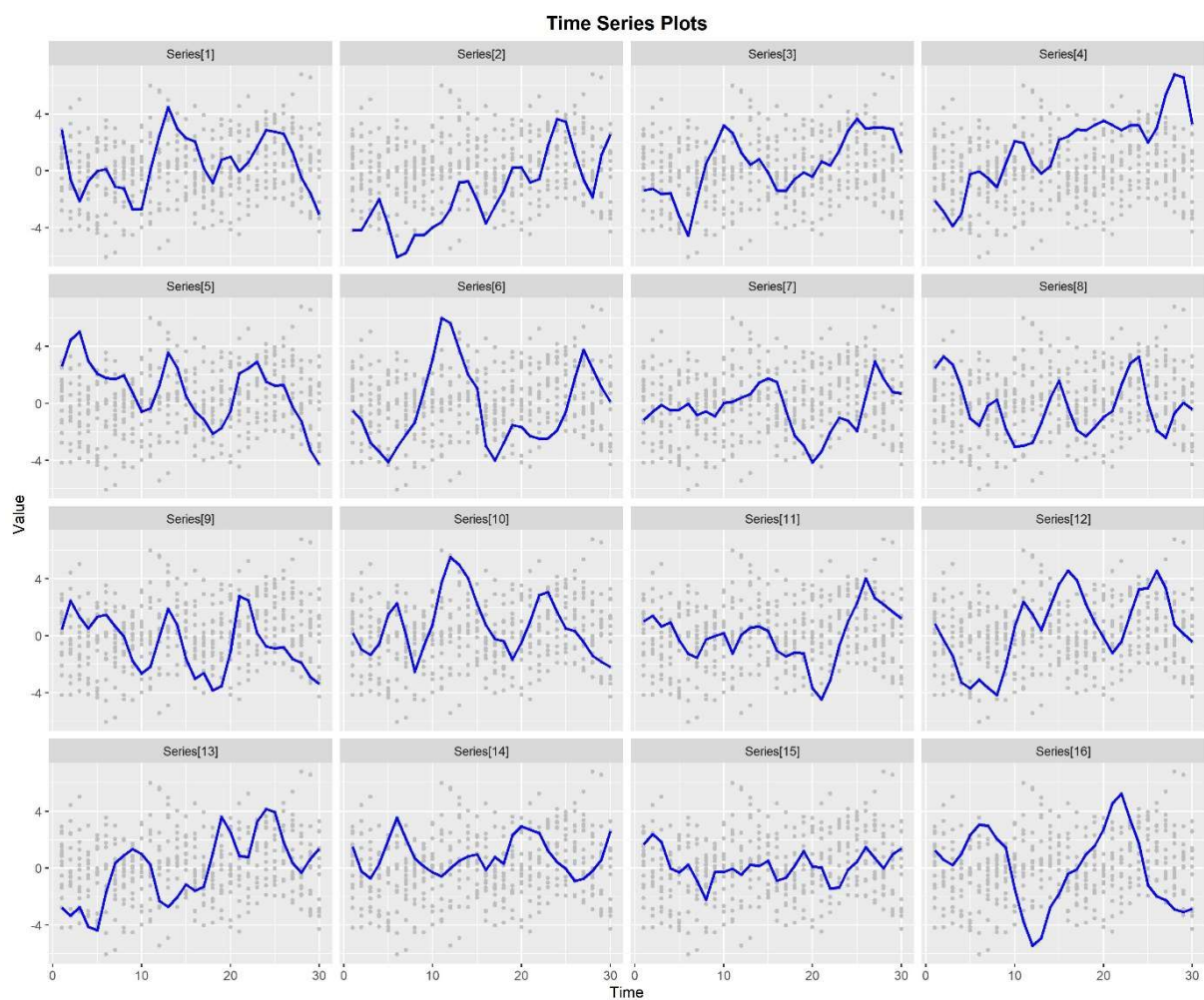

**Figure 1:** Output of the `plot` function for `n = 16` vectors generated with `rGARMA`
(time-series vectors were generated from a GARMA(2,2) model)

To illustrate the conditioning syntax in our package we will also show another example where we generate time-series vectors from the same model, but with a set of conditioning values. We again use the **rGARMA** function to generate $n = 16$ random time-series vectors of length $m = 30$ from this model, but we impose conditioning value $y_1 = -4$, $y_{12} = 0$ and $y_{30} = 4$. The code is shown below and the resulting scatterplot matrix is shown in Figure 2 below; as can be seen, each time-series goes through the specified conditioning points.

```
#Set the conditional values
CONDVALS     <- rep(NA, 30)
CONDVALS[1]  <- -4
CONDVALS[12] <-  0
CONDVALS[30] <-  4

#Generate and plot random time-series from the GARMA distribution
SERIES.COND <- rGARMA(n = 16, m = 30, ar = AR, ma = MA,
                      condvals = CONDVALS)
plot(SERIES.COND)
```

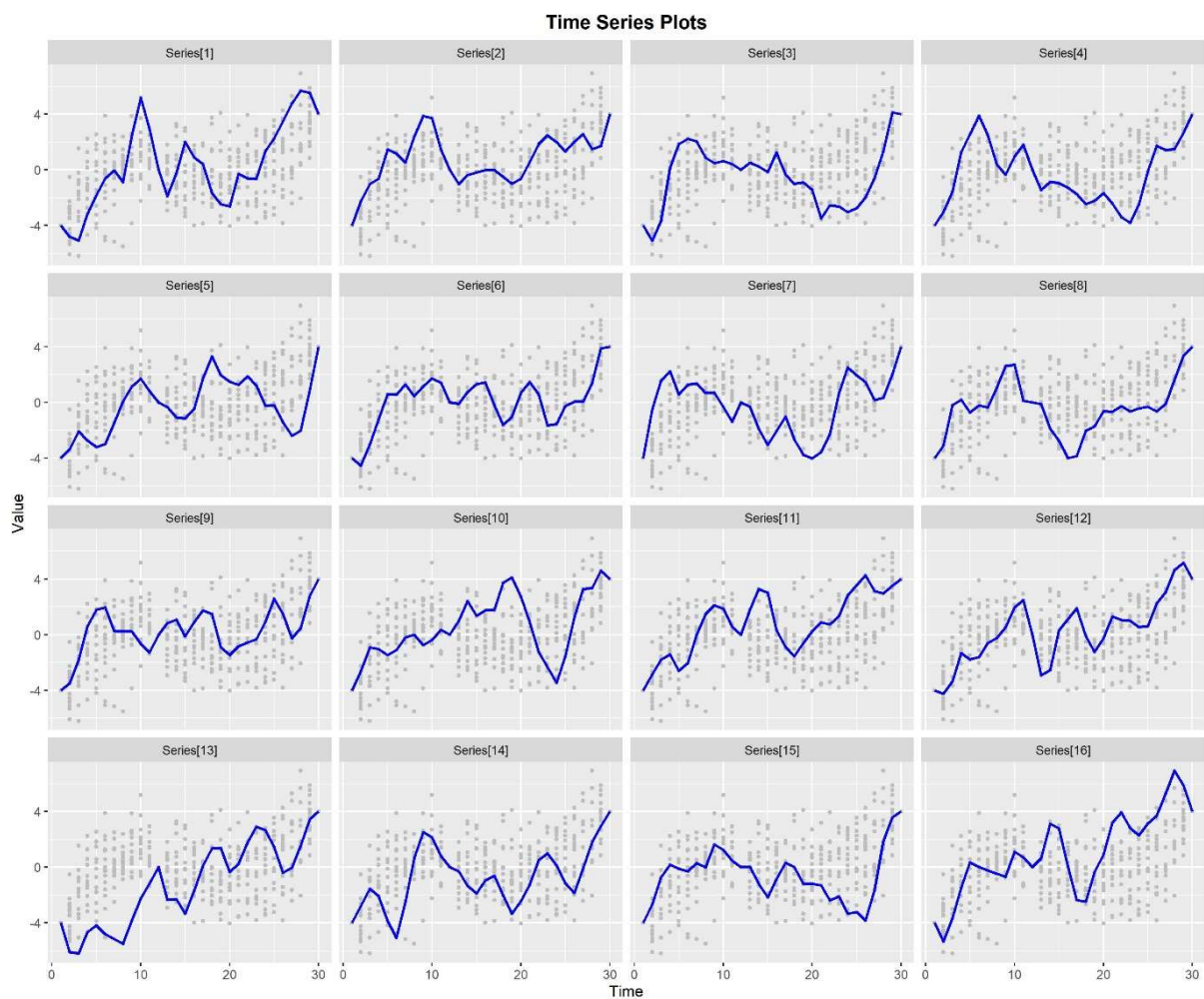

**Figure 2:** Output of the **plot** function for **n = 16** vectors generated with **rGARMA**
(time-series vectors were generated from a GARMA(2,2) model with conditioning values)

We can see the conditioning values in the above figure; the background dots show all the values generated at each time index. As specified in the **condvals** input, every time-series vector in the figure starts at the value $y_1 = -4$ and then goes through the value $y_{12} = 0$ and ends at the value $y_{30} = 4$. As a result, each generated time-series vector now has a general "upward" trajectory. The code to impose this conditioning requirement was extremely simple; all we did was to create a vector of conditioning values with **NA** values on the non-conditional values. This makes the function highly user-friendly for producing conditional random vectors.

If we already have a matrix of time-series vectors we can use the **dGARMA** or **pGARMA** function to get the density or cumulative probability of these vectors from a specified GARMA model. In the code below we generate the density and cumulative probability values for our generated time-series vectors from the model that was used to generate these vectors. As is standard for density and cumulative probability functions in **R**, both functions are "vectorised" so that they can take in a matrix with rows giving the individual time-series vectors. We input our **SERIES** matrix which contains $n = 16$ rows giving time-series vectors from the GARMA model, which means that the output is a vector with 16 elements giving the density or cumulative probability for each time-series in the input matrix.

```
#Compute the density of the GARMA output
(DENSITY <- dGARMA(SERIES, ar = AR, ma = MA))

    Series[1]     Series[2]     Series[3]     Series[4]     Series[5]
 1.129149e-21 5.397295e-22 4.160202e-19 2.568389e-20 9.069298e-20
    Series[6]     Series[7]     Series[8]     Series[9]    Series[10]
 4.296245e-21 4.254306e-18 9.037405e-19 1.248155e-20 7.739927e-20
   Series[11]    Series[12]    Series[13]    Series[14]    Series[15]
 5.535573e-19 6.446935e-20 2.566024e-20 2.693398e-18 4.643742e-19
   Series[16]
 5.323142e-19

#Compute the cumulative probability of the GARMA output
(PROBS <- pGARMA(SERIES, ar = AR, ma = MA))

    Series[1]     Series[2]     Series[3]     Series[4]     Series[5]
 6.150840e-05 1.798479e-09 8.705554e-05 1.059079e-03 1.583684e-04
    Series[6]     Series[7]     Series[8]     Series[9]    Series[10]
 1.541055e-06 1.607233e-05 1.251148e-05 1.384605e-06 1.477541e-04
   Series[11]    Series[12]    Series[13]    Series[14]    Series[15]
 4.456023e-05 5.011268e-05 6.503880e-06 2.430163e-03 3.408341e-04
   Series[16]
 3.987137e-06
```

In addition to the probability functions for the GARMA model, the package allows the user to compute the auto-covariance/correlation function using the **ARMA.acf** function or compute the auto-covariance/correlation matrix using the **ARMA.var** function. These two functions are variations of the **ARMAacf** function in the **stats** package in base **R** (R Core Team 2019), which uses a recursive method based on Section 3.3 of Brockwell and Davis (1991). The inputs for these functions are shown below in Table 3. The output of the **ARMA.acf** function is a vector of auto-covariance/correlation values indexed by lag values (starting at zero), and the output of the **ARMA.var** function is a matrix of auto-covariance/correlation value indexed by time values (starting at one).

| Argument | Type | Description |
|---|---|---|
| **n** | **Integer** (positive) | An integer specifying the number of time values (the maximum lag is one less than this value). |
| **condvals*†** | **Numeric** (vector) | Numeric conditioning values for generating a random time-series. This input must either be a single value **NA** (indicating that there are no conditioning values) or it must be a numeric vector with length **n**. Each numeric value in the vector is taken as a conditioning value and each **NA** shows that there is no conditioning value for that element (i.e., that element is to be randomly generated). |
| **ar*** | **Numeric** (vector) | The vector of auto-regressive coefficients for the GARMA distribution. |
| **ma*** | **Numeric** (vector) | The vector of moving-average coefficients for the GARMA distribution. |
| **corr*** | **Logical** (value) | Logical value specifying whether the user wants the correlation values; if **TRUE** then the output is the auto-correlation rather than the auto-covariance. |
| \* Non-mandatory argument; default values are given in the function. ||||
| † This input is used in the **ARMA.var** function but not in the **ARMA.acf** function. ||||

**Table 3:** Arguments for the **ARMA.acf** and **ARMA.var** functions

In the code below we use the **ARMA.acf** function to show the auto-correlation values up to lag $\ell = 5$ in a GARMA model using the above auto-regressive and moving-average polynomials. The corresponding auto-correlation matrix from the **ARMA.var** function is a $6 \times 6$ matrix with these correlation values on the appropriate diagonals; for brevity we do not show the output of this function here.

```
#Compute the auto-correlation function
ARMA.acf(n = 6, ar = AR, ma = MA, corr = TRUE)

    Lag[0]     Lag[1]     Lag[2]     Lag[3]     Lag[4]     Lag[5]
1.00000000 0.83519207 0.52763321 0.25506815 0.09852788 0.02780867
```

# 4. Other functions in the package

In addition to the probability functions for the GARMA model, the `ts.extend` package contains functions to compute the Fourier intensity of a time-series vector, and test the exchangeability of a time-series vector using a permutation-spectrum test (which we discuss soon). The first of these is done using the `intensity` function. This function takes a time-series vector and computes the corresponding vector of Fourier intensity values (i.e., the norms of the complex values in the discrete Fourier transform). The inputs to this function are shown in Table 4. The function allows both real and complex time-series vectors as inputs.

| Argument | Type | Description |
|---|---|---|
| `x` | **Numeric/ Complex** (vector/matrix) | An input vector or matrix (real or complex). If the input is a vector it is treated as a single time-series. If the input is a matrix, each row of the matrix is treated as a time-series. |
| `centred/ centered*` | **Logical** (value) | Logical value indicating whether the time-series vector should be centred before computing the intensity; if `TRUE` then the time-series vector is centred prior to computation, which means that the intensity at the zero frequency is zero. |
| `scaled*` | **Logical** (value) | Logical value indicating whether the intensity should be scaled; if `TRUE` then the output vector gives the scaled intensity. |
| `nyquist*` | **Logical** (value) | Logical value indicating whether to restrict intensity of real time-series vectors to the Nyquist range; if `TRUE` then the intensity vector for a real time-series vector is truncated to restrict the frequencies from zero to one-half. |
| `*` Non-mandatory argument; default values are given in the function. | | |

**Table 4:** Arguments for the `intensity` function

The `intensity` function allows the user to centre the input vector prior to computing the intensity (which gives zero intensity at the zero frequency) and also allows the user to compute the scaled intensity instead of the raw intensity; the former subtracts the sample mean and the latter scales the time-series vector by dividing by its sample standard deviation.[6] For real time-series vectors the intensity vector is symmetric around the Nyquist frequency of one-half, so that the latter part of the vector gives no information about the time-series. Consequently, the function gives the user the option to truncate the output vector to remove this latter part and restrict attention to the Nyquist range.[7]

---

[6] If the time-series vector was not centred then the degrees-of-freedom is equal to the length of the vector; if the time-series vector was centred then the degrees-of-freedom is equal to the length of the vector minus one. The sum of squares of the scaled intensity values is equal to the degrees of freedom.

[7] To avoid any confusion, it is worth noting that if this option is selected then the output is truncated so that it is not the full intensity vector for the DFT. This means that the scaled intensity values will not necessarily sum to an integer, as might be expected.

In the code below we generate and plot the Fourier intensity of our first time-series vector using the `intensity` function. The resulting intensity plot is shown in Figure 3 below; by default the function centres and scales the time-series vector, so the figure shows the scaled intensity on the vertical axis. As is unsurprising for our model, the auto-correlation at lower lag values leads to higher intensity at the lower frequencies. This particular time-series vector oscillates

```
#Show the intensity of a time-series vector
SERIES1 <- SERIES[1,]
INT     <- intensity(SERIES1)
c(INT)

 Freq[0/30]  Freq[1/30]  Freq[2/30]  Freq[3/30]  Freq[4/30]
  0.0000000   1.8053526   1.1445994   2.1577709   0.5116496
 Freq[5/30]  Freq[6/30]  Freq[7/30]  Freq[8/30]  Freq[9/30]
  1.3657857   0.6216972   0.2911577   1.1053203   0.2946183
Freq[10/30] Freq[11/30] Freq[12/30] Freq[13/30] Freq[14/30]
  0.6730260   0.2545078   0.5882391   0.6033748   0.2780761
Freq[15/30]
  0.3535510

plot(INT)
```

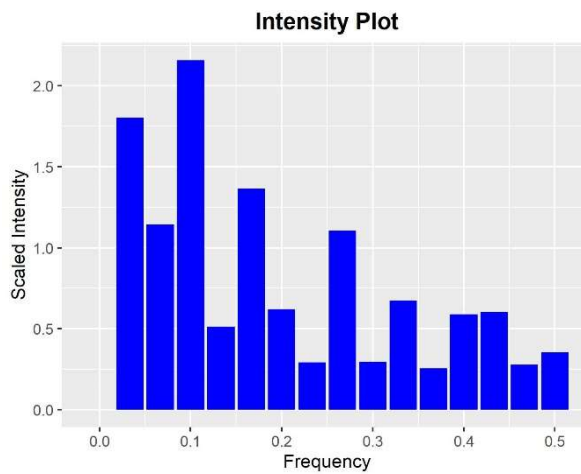

**Figure 3:** Output of the `plot` function intensity generated with `intensity`

The remaining function in the package is the `spectrum.test` function, which implements the permutation-spectrum test in O'Neill (2021). This is a test comparing the null hypothesis of exchangeability against the alternative hypothesis that there is a signal in the data.[8] The test uses the maximum signal intensity as a test statistic and compares this to its null distribution under exchangeability. The null distribution is computed using simulated permutations of the input time-series vector, leaning to a generated vector of maximum spectrum intensity values for the permutations (hence the name of the test). The inputs to this function are shown in Table 5. The function allows both real and complex time-series vectors as inputs, and also allows the user to specify the number of simulations to approximate the null distribution.

---

[8] This can either be a periodic signal or a quasi-periodic signal determined by non-zero auto-correlation in the time-series process; see O'Neill (2021) (Section 1) for further details.

| Argument | Type | Description |
|---|---|---|
| `x` | Numeric/ Complex (vector) | An input time-series vector (real or complex). |
| `sims*` | Integer (positive) | The number of simulations to use for the test (default is one-million simulations). |
| `progress*` | Logical (value) | Logical value indicating whether the user wants a progress bar; if **TRUE** then the computation will be accompanied by a progress bar that tracks the number of simulations completed. |
| **\*** Non-mandatory argument; default values are given in the function. | | |

**Table 5:** Arguments for the `spectrum.test` function

To work effectively, the permutation-spectrum test requires a large number of simulations that each involve taking a permutation of the input time-series vector and computing its maximum spectrum intensity. For long input vectors this can take a substantial computation time. Thus, to facilitate the user of the `spectrum.test` function, there is a logical input that gives the user the option to have a progress bar to keep track of progress. By default this progress bar is present in the function, but it can be turned off using the logical input `progress` if it becomes a nuisance.

In the code below we perform the permutation-spectrum test on our first generated time-series vector using the `spectrum.test` function. The resulting plot is shown in Figure 4 below. This figure shows the intensity plot on the left and the simulated null distribution on the right; the maximum intensity value is shown as a red dot in both plots. As can be seen from the output, in this particular case we have a p-value of 11.33% so there is insufficient evidence to reject the hypothesis of exchangeability.

```
#Show the intensity of a time-series vector
TEST <- spectrum.test(SERIES1)

===============================================================

TEST
```

```
           Permutation-Spectrum Test

data:   real time-series vector SERIES1 with 30 values
maximum scaled intensity = 2.1578, p-value = 0.1133
alternative hypothesis: distribution of time-series vector is not
exchangeable (at least one periodic signal is present)
```

**plot(TEST)**

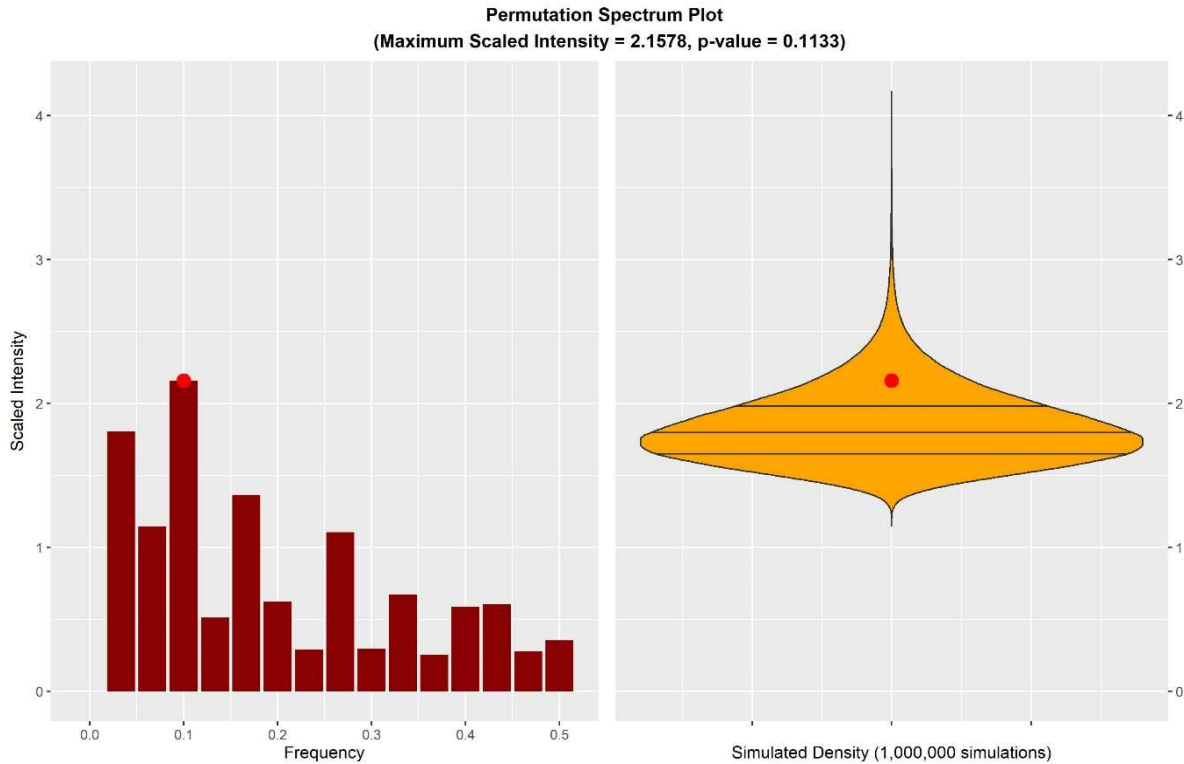

**Figure 4:** Output of the `plot` function intensity generated with `spectrum.test`

Details on the methodology and interpretation of the permutation-spectrum test can be found in O'Neill (2021). The general idea can be gleaned from the test output and plot. In the above case the maximum scaled intensity of the time-series is somewhat in the upper tail of the null distribution, but it is not particularly extreme. The test output gives a p-value of $p = 0.1133$ which means that random permutations of this time-series vector would give a maximum scaled intensity value that is at least as large as the observed value a bit over eleven percent of the time. In this example there is no evidence of a periodic signal in the time-series.

The permutation-spectrum test is useful in a variety of settings where an analyst wishes to test a time-series vector for the presence of a periodic signal. It is particularly useful as a diagnostic test in periodic regression models and other time-series models, where it is desired to test a residual time-series vector to see if all periodic signals have been removed. The test could be used to supplement other diagnostic tests applied to ARMA models or time series models more generally (see e.g., Francq, Roy and Zakoïan 2005).

## 5. Overview

The `ts.extend` package provides probability functions for the stationary Gaussian ARMA model, to allow the user to compute the density or distribution function, or generate random vectors from the distribution. This can be done marginally or conditionally, using simple syntax. The package also includes functions to compute the spectral intensity of a time-series vector and to implement the permutation-spectrum test for exchangeability of the vector. The package has custom plotting functions for its outputs to allow the user to easily produce plots showing time-series vectors, their spectral intensity, or the results of the permutation-spectrum test. The package is designed to assist in the simulation and analysis of time-series data, particularly when dealing with Gaussian ARMA models.

## References

ARMSTRONG, W. (2018) fts: R interface to 'tslib' (a time series library in C++). *R Package*, Version 0.9.9.2. https://CRAN.R-project.org/package=fts

BOX, G.E.P. AND JENKINS, G. (1970) *Time Series Analysis: Forecasting and Control*. Holden Day: San Fransisco.

BROCKWELL, P.J. AND DAVIS, R.A. (1991) Time Series: Theory and Methods (Second Edition). Springer: New York.

JAMES, D., HORNIK, K., GROTHENDIECK, G. AND R CORE TEAM. (2020) chron: chronological objects which can handle dates and times. *R Package*, Version 2.3-54. https://CRAN.R-project.org/package=chron

FRANCQ, C., ROY, R. AND ZAKOÏAN, J-M. (2005) Diagnostic checking in ARMA models with uncorrelated errors. *Journal of the American Statistical Association* **100(470)**, pp. 532-544.

FULLER, W.A. (1996) *Introduction to Statistical Time Series (Second Edition)*. John Wiley and Sons: New York.

HOLAN, S.H., LUND, R. AND DAVIS, G. (2000) The ARMA alphabet soup: a tour of ARMA model variants. *Statistics Surveys* **4**, pp. 232-274.

HYNDMAN, R., ATHANASOPOULOS, G., BERGMEIR, C., CACERES, G., CHHAY, L., O'HARA-WILD, M., PETROPOLOUS, F., RAZBASH, S., WANG, E., YASMEEN, F., R CORE TEAM, IHAKA, R., REID, D., SHAUB, D., TANG, Y. AND ZHOU, Z. (2021) forecast: forecasting functions for time series and linear models. *R Package*, Version 8.15. https://CRAN.R-project.org/package=forecast

Kourentzes, N., Svetunkov, I. and Schaer, O. (2020) tsutils: Time Series Exploration, Modelling and Forecasting. *R Package*, Version 0.9.2. https://CRAN.R-project.org/package=tsutils

Krispin, R. (2020) TSstudio: functions for time series analysis and forecasting. *R Package*, Version 0.1.6. https://CRAN.R-project.org/package=TSstudio

O'Neill, B. (2020) ts.extend: stationary Gaussian ARMA processes and other time-series utilities. *R package*, Version 0.1.1. https://CRAN.R-project.org/package=ts.extend

O'Neill, B. (2021) The permutation-spectrum test: identifying periodic signals using the maximum Fourier intensity. *ArXiv preprint*, arXiv:2109.05798.

Priestley, M.B. (1981) *Spectral Analysis and Time Series*. Academic Press: London.

R Core Team (2019) R: A language and environment for statistical computing. R Foundation for Statistical Computing: Vienna. https://www.R-project.org/

Ryan, J.A., Ulrich, J.M, Bennett, R. and Joy, C. (2020) xts: extensible time series. *R Package*, Version 0.12.1. https://CRAN.R-project.org/package=xts

Slutsky, E. (1937) The summation of random causes as the source of cyclic processes. *Econometrica* **5(2)**, pp. 105-146.

Spinu, V., Grolemund, G., Wickham, H., Lyttle, I., Costigan, I., Law, J., Mitarotonda, D., Larmarange, J., Boiser, J., Lee, C.H. and Google Inc. (2021) lubridate: make dealing with dates a little easier. *R Package*, Version 1.7.10. https://CRAN.R-project.org/package=lubridate

Trapletti, A., Hornik, K. and LeBaron, B. (2020) tseries: time series analysis and computational finance. *R Package*, Version 0.10-48. https://CRAN.R-project.org/package=tseries

Wold, H. (1954) *A Study in the Analysis of Stationary Time Series*. Almquist and Wiksell: Stockholm.

Wuertz, D., Setz, T., Chalabi, Y. and Maechler, M. (2018) timeDate: Rmetrics — chronological and calendar objects. *R Package*, Version 3043.102. https://CRAN.R-project.org/package=timeDate

Yaglom, A.M. (1987) *Correlation Theory of Stationary and Related Random Functions (Volume I)*. Springer: New York.

Yule, G.U. (1926) Why do we sometimes get nonsense-correlations between time series? A study in sampling and the nature of time series. *Journal of the Royal Statistical Society* **89(1)**, pp. 1-63.

Zeileis, A., Grothendieck, G., Ryan, J.A., Ulrich, J.M. and Andrews, F. (2021) zoo: S3 infrastructure for regular and irregular time series (z's ordered observations). *R Package*, Version 1.8-9. https://CRAN.R-project.org/package=zoo